\documentclass[12pt,epsf]{article}
\usepackage[dvips]{graphicx}

\baselineskip=\normalbaselineskip

\newcommand{\resection}[1]{\setcounter{equation}{0}\section{#1}}
\renewcommand{\theequation}{\thesection.\arabic{equation}}
\renewcommand{\thefootnote}{\fnsymbol{footnote}}
\newcommand{\bel}[1]{\begin{equation}\label{#1}}
\newcommand{\bal}[1]{\begin{eqnarray}\label{#1}}

\newcommand{\be}{\begin{equation}}
\newcommand{\ee}{\end{equation}}
\newcommand{\ba}{\begin{eqnarray}}
\newcommand{\ea}{\end{eqnarray}}
\newcommand{\nn}{\nonumber \\}

\newcommand{\tr}{{\rm tr}}
\newcommand{\bR}{{\bf R}}

\newcommand{\eq}[1]{(\ref{#1})}
\newcommand{\cH}{{\cal H}}
\newcommand{\cL}{{\cal L}}
\newcommand{\cP}{{\cal P}}
\newcommand{\Sloc}{{S_{\rm loc}}}
\newcommand{\cLloc}{{{\cal L}_{\rm loc}}}
\newcommand{\cJ}{{\cal J}}

\newcommand{\cW}{{\cal W}}
\newcommand{\gh}{\widehat{g}}
\newcommand{\hh}{\widehat{h}}
\newcommand{\Dh}{\widehat{D}}
\newcommand{\Gb}{\overline{G}}
\newcommand{\phib}{\bar{\phi}}
\newcommand{\Pib}{\overline{\Pi}}

\newcommand{\bG}{\mbox{\boldmath $G$}}
\renewcommand{\bR}{\mbox{\boldmath $R$}}
\ifx\TwoupWrites\UnDeFiNeD\else\target{\magstepminus1}{11.3in}{8.27in}
\source{\magstep0}{7.5in}{11.69in}\fi
\newfont{\fourteencp}{cmcsc10 scaled\magstep2}
\newfont{\titlefont}{cmbx10 scaled\magstep3}
\newfont{\authorfont}{cmcsc10 scaled\magstep1}
\newfont{\fourteenmib}{cmmib10 scaled\magstep2}
\skewchar\fourteenmib='177
\newfont{\elevenmib}{cmmib10 scaled\magstephalf}
\skewchar\elevenmib='177
\makeatletter
\newcommand\nonsequentialeqnum{
\@addtoreset{equation}{section}
\def\theequation{\arabic{section}.\arabic{equation}}}
\newif\ifp@bblock \p@bblocktrue
\newcommand\nopubblock{\p@bblockfalse}
\newcommand\topspace{\hrule height 0pt depth 0pt \vskip}
\newcommand\p@bblock{\begingroup \tabskip=\hsize minus \hsize
\baselineskip=1.5\ht\strutbox \topspace-2\baselineskip
\halign to\hsize{\strut ##\hfil\tabskip=0pt\crcr
\the\Pubnum\crcr\the\date\crcr}\endgroup}

\renewcommand\titlepage{\ifx\TwoupWrites\UnDeFiNeD\null
\vspace{-1.7cm}\fi
\vskip0.6cm
\ifp@bblock\p@bblock \else\hrule height 0pt \relax \fi}
\makeatother
\newtoks\date
\newtoks\Pubnum
\newtoks\pubnum
\Pubnum={
~ \crcr
~ \crcr
YITP-00-37 \crcr
{\tt hep-th/0007062} \crcr
{}July 2000
}
\newcommand{\frontpageskip}{\vspace{12pt plus .5fil minus 2pt}}
\renewcommand{\title}[1]{\frontpageskip
\begin{center}{\titlefont #1}\end{center}\par}
\renewcommand{\author}[1]{\frontpageskip\par\begin{center}
{\authorfont #1}\end{center}
\nobreak
}

\renewcommand{\thanks}[1]{\footnote{#1}}
\renewcommand{\abstract}{\par\frontpageskip\centerline{
\fourteencp Abstract}
\vspace{8pt plus 3pt minus 3pt}}

\begin{document}

\begin{titlepage}

\vspace*{\fill}
\begin{center}
{\Large{\bf A Note on the Weyl Anomaly\\ 
in the Holographic Renormalization Group }} \\
\vfill
{\sc Masafumi Fukuma
\footnote{e-mail: {\tt fukuma@yukawa.kyoto-u.ac.jp}},
\sc So Matsuura
\footnote{e-mail: {\tt matsu@yukawa.kyoto-u.jp}}
and
{\sc Tadakatsu Sakai}
\footnote{e-mail: {\tt tsakai@yukawa.kyoto-u.ac.jp}}}\\[2em]

{\sl Yukawa Institute for Theoretical Physics,\\
      Kyoto University, Kyoto 606-8502, Japan } \\

\vfill
ABSTRACT
\end{center}
\begin{quote}
We give a prescription for calculating the holographic Weyl anomaly 
in arbitrary dimension within the framework based 
on the Hamilton-Jacobi equation proposed by 
de Boer, Verlinde and Verlinde. 
A few sample calculations are made and shown to 
reproduce the results that are obtained to this time 
with a different method. 
We further discuss continuum limits, 
and argue that the holographic renormalization group 
may describe the renormalized trajectory in the parameter space. 
We also clarify the relationship of the present formalism 
to the analysis carried out by Henningson and Skenderis. 
\end{quote}
\vfill
\end{titlepage}

%
\renewcommand{\thefootnote}{\arabic{footnote}}
\setcounter{footnote}{0}
\addtocounter{page}{1}%

\resection{Introduction}
The AdS/CFT correspondence \cite{MGKPW} 
(for a review see Ref.\ \cite{review}) 
states that a gravitational theory 
on the $(d+1)$-dimensional anti-de-Sitter space (AdS$_{d+1}$) has a dual 
description in terms of a conformal field theory 
on the $d$-dimensional boundary. 
One of the most significant aspects of the AdS/CFT correspondence 
is that it can further give us a framework to study the 
renormalization group (RG) structure of the boundary field theories 
\cite{E.T.A}\cite{AG}\cite{HRG}\cite{GPPZ}\cite{GPPZ-2}\cite{PS}%
\cite{BPPZ}\cite{ST}\cite{DFGK}. 
In this scheme of the  ``holographic RG,'' 
the extra radial coordinate in the bulk is 
regarded as parametrizing the RG flow of the dual boundary field theory, 
and the evolution of bulk fields along the radial direction 
is considered as describing the RG flow of the coupling constants 
in the boundary field theory.

In Ref.\ \cite{dVV}, de Boer, Verlinde and Verlinde proposed 
the formulation of the holographic RG based on 
the Hamilton-Jacobi equation. 
They showed, by investigating five-dimensional gravity with scalar
fields, that the Callan-Symanzik equation of the four-dimensional 
boundary theory actually arises from the holographic RG.
They also calculated the Weyl anomaly in four dimensions 
and found that the result agrees with those given 
in Ref.\ \cite{HS;weyl}
(see Ref.\ \cite{Duff;Weyl} for a review of the Weyl anomaly). 
The extension of their analysis to a system including gauge fields 
is discussed in Ref.\ \cite{vector}.

The first main aim of the present note is to give 
a prescription for calculating the Weyl anomaly in arbitrary dimension,
within the framework based on the Hamilton-Jacobi equation. 
This prescription is actually a simple generalization of the 
algorithm given in Ref.\ \cite{dVV} for the four-dimensional case.
Here we carry out a few sample calculations to affirm its correctness.

Second, we give discussion on continuum limits, 
and show that when bare couplings are tuned such that 
they are on the classical trajectories passing through the corresponding 
renormalized couplings, 
both the bare and renormalized couplings 
satisfy an RG equation of the same functional form. 
This fact strongly suggests that the holographic RG may directly 
describe the so-called renormalized trajectory \cite{wilson-kogut}  
in the parameter space.

Finally, we discuss the relationship among various renormalizations 
adopted in the literature on the holographic RG. 
In particular, we give a detailed analysis of the relationship 
between the analysis based on the Hamilton-Jacobi equation 
and that carried out by Henningson and Skenderis \cite{HS;weyl}.

The organization of this note is as follows. 
In \S 2, we give a review of the flow equation that is 
obtained from the Hamilton-Jacobi equation \cite{dVV}. 
In \S 3, we describe a prescription for solving the flow equation 
and make sample calculations of the Weyl anomaly 
in four and six dimensions. 
The results are found to agree with those given 
in Ref.\ \cite{HS;weyl}. 
In \S 4, we explore the continuum limits of the boundary field theory 
in the context of the holographic RG. 
In \S 5, we investigate the relationship among 
various renormalizations. 
In particular, we give a detailed discussion 
of the relation between the present analysis and 
that given in Ref.\ \cite{HS;weyl}. 
Section 6 is devoted to conclusions. 
The appendices are meant to make this note as self-contained as possible.

%
\resection{Hamilton-Jacobi constraint and the flow equation}

In this section, we briefly review the formulation of the holographic
RG based on the Hamilton-Jacobi equation \cite{dVV},  
with the purpose of fixing our convention.

We start by recalling the Euclidean ADM decomposition 
that parametrizes a $(d+1)$-dimensional metric as
\ba
 ds^2&=&\bG_{M\!N}\,dX^M dX^N \nn
 &=&N(x,r)^2 dr^2+G_{\mu\nu}(x,r)\Bigl(dx^{\mu}+\lambda^\mu(x,r)dr\Bigr) 
 \Bigl(dx^{\nu}+\lambda^{\nu}(x,r)dr\Bigr).
\label{metric;dvv}
\ea
Here $X^M=(x^\mu,r)$ with $\mu,\nu=1,2,\cdots,d$, 
and $N$ and $\lambda^\mu$ are the lapse and the shift function, 
respectively. 
The signature of the metric $G_{\mu\nu}$ is taken to be $(+\cdots +)$. 
As we discussed in the Introduction, the Euclidean time $r$ 
is identified with 
the RG parameter of the $d$-dimensional boundary theory, 
and the evolution of bulk fields in $r$ 
is identified with the RG flow of the coupling constants 
of the boundary theory. 
In the following discussion, we exclusively consider scalar fields 
as such bulk fields.

The Einstein-Hilbert action with bulk scalars $\phi^i(x,r)$ 
on a $(d+1)$-dimensional manifold $M_{d+1}$ 
with boundary $\Sigma_d=\partial M_{d+1}$ is given by 
\begin{eqnarray}
 &&S_{d+1}[\bG_{M\!N}(x,r),\phi^i(x,r)] \nn
 &&~~~~~=\int_{M_{d+1}} d^{d+1} X \sqrt{\bG} \left( V( \phi ) 
  -\bR+{1 \over 2}\,L_{ij}( \phi )\, 
  \bG^{M\!N}\,\partial_M \phi^i\,\partial_N \phi^j \right)
  -2\int_{\Sigma_d} d^d x \,\sqrt{G}\,K\,,\nn
 && \label{bulk}
\end{eqnarray}
which is expressed in the ADM parametrization as 
\begin{eqnarray}
 &&S_{d+1}[G_{\mu\nu}(x,r),\phi^i(x,r),N(x,r),\lambda^\mu(x,r)]\nn
 &&~~~=\int d^dx\,dr \,\sqrt{G}\, \Bigl[\, N 
  \left( V(\phi)-R+K_{\mu\nu}K^{\mu\nu}-K^2 \right) \nonumber \\
 &&~~~~~  
   +\,{1 \over 2N}\,L_{ij}(\phi) \left( 
  \left(\dot{\phi}^i -\lambda^{\mu}\partial_{\mu}\phi^i\right) 
  \left(\dot{\phi}^j -\lambda^{\mu}\partial_{\mu}\phi^j\right) 
  +N^2 G^{\mu\nu}\partial_{\mu}\phi^i\partial_{\nu}\phi^j 
  \right ) \,\Bigr] \nn
 &&~~~\equiv \int d^d x \,dr \,\sqrt{G} \,\cL_{d+1}[G,\phi,N,\lambda],
\label{eh;off-shell}
\end{eqnarray}
where $ \cdot=\partial/\partial r$. 
Here $R$ and $\nabla_{\mu}$ are the scalar curvature and the covariant
derivative with respect to $G_{\mu\nu}$, respectively, 
and $K_{\mu\nu}$ is the extrinsic curvature on $\Sigma_d$ given by 
\begin{equation}
 K_{\mu\nu}={1 \over 2N}
  \left(\dot{G}_{\mu\nu}-\nabla_{\mu}\lambda_{\nu}
  -\nabla_{\nu}\lambda_{\mu}\right),\qquad
 K=G^{\mu\nu}\,K_{\mu\nu}. 
\end{equation}
The boundary term in Eq.\ (\ref{bulk}) needs to be introduced 
to ensure that the Dirichlet 
boundary conditions can be imposed on the system consistently \cite{GH}. 
In fact, the second derivative in $r$ appearing in the first term 
of Eq.\ (\ref{bulk}) can be written as a total derivative 
and canceled with the boundary term.

As far as classical solutions are concerned, the action \eq{eh;off-shell}
is equivalent to the following one in first-order form: 
\begin{eqnarray}
 S_{d+1}[G_{\mu\nu},\phi^i,\Pi^{\mu\nu},\Pi_i, N,\lambda^\mu]
 \equiv \int d^d x \,dr \,\sqrt{G} 
  \left[\,\Pi^{\mu\nu}\dot{G}_{\mu\nu}+\Pi_i\dot{\phi}^i
  +N\cH+\lambda_{\mu}\cP^{\mu}\,\right], \nonumber \\
 \label{first}
\end{eqnarray}
with 
\ba
 \cH &\equiv& {1 \over d-1}\left(\Pi_{\mu}^{\mu}\right)^2-\Pi_{\mu\nu}^2
 -{1 \over 2}\,L^{ij}(\phi)\,\Pi_i\,\Pi_j+V(\phi)-R
 +{1 \over 2}\,L_{ij}(\phi)\,G^{\mu\nu}\,
  \partial_{\mu}\phi^i\,\partial_{\nu}\phi^j, 
  \nn
 \cP^{\mu}&\equiv& 2\,\nabla_{\nu}\Pi^{\mu\nu}-\Pi_i\,\nabla^{\mu}\phi^i. 
\ea
In fact, the equations of motion for $\Pi^{\mu\nu}$ and 
$\Pi_i$ give the relations 
\begin{equation}
 \Pi^{\mu\nu}
  =K^{\mu\nu}-G^{\mu\nu}K, 
  \qquad \Pi_i
  ={1 \over N} L_{ij}(\phi) \left( \dot{\phi}^j -
  \lambda^{\mu}\,\partial_{\mu}\phi^j \right), 
\label{Pi;gphi}
\end{equation}
and by substituting this expression into Eq.\ \eq{first}, 
we can obtain \eq{eh;off-shell}. 
Here $N$ and $\lambda^\mu$ simply behave as 
Lagrange multipliers, giving
the Hamiltonian and momentum constraints: 
\begin{eqnarray}
 {1 \over \sqrt{G}}\,{\delta S_{d+1} \over \delta N}&=&\cH~=~0,
 \label{h;constraint}\\
 {1 \over \sqrt{G}}\,{\delta S_{d+1} \over \delta \lambda_{\mu}}
  &=&\cP^{\mu}~=~0.
 \label{m;constraint}
\end{eqnarray}
Note that these constraints generate reparametrizations
of the form $r \rightarrow r+\delta r(x),\,
x^\mu \rightarrow x^\mu+\delta x^\mu(x)$ 
for systems on an ``equal time slice'' $\Sigma_d$ ($r={\rm const}$).
One can easily show that they are of the first class 
under the canonical Poisson brackets for $G_{\mu\nu},\Pi^{\mu\nu},
\phi^i$ and $\Pi_i$. 
Thus, up to gauge equivalent configurations generated by 
$\cH$ and $\cP^\mu$, 
the $r$-evolution of the bulk fields 
is uniquely determined, being independent of the values 
of the Lagrange multiplier $N$ and $\lambda^\mu$. 
In the following discussion, we work in the ``temporal gauge,'' 
$N=1,~\lambda^{\mu}=0$.


Let $\Gb_{\mu\nu}(x,r;G(x),r_0)$ and $\phib^i(x,r;\phi(x),r_0)$
be the classical solutions of the bulk action 
with the boundary conditions\footnote{
One generally needs two boundary conditions for each field, 
since the equation of motion is a second-order differential equation
in $r$. 
Here, one of the two is assumed to be already fixed 
by demanding the regular behavior of the classical solutions 
inside $M_{d+1}$ ($r\rightarrow +\infty$) \cite{MGKPW} 
(see also Ref.\ \cite{GL}).
}
\begin{equation}
 \Gb_{\mu\nu}(x,r\!=\!r_0)=G_{\mu\nu}(x),\qquad
 \phib^i(x,r\!=\!r_0)=\phi^i(x).
 \end{equation}
We also define $\Pib^{\mu\nu}(x,r)$ and $\Pib_i(x,r)$ to be the classical 
solutions of $\Pi^{\mu\nu}(x,r)$ and $\Pi_i(x,r)$, respectively. 
We then define the on-shell action that is obtained 
as a functional of the boundary values, $G_{\mu\nu}(x)$ and $\phi^i(x)$, 
by substituting these classical solutions into the bulk action:
\begin{eqnarray}
 &&S[G_{\mu\nu}(x),\phi(x),r_0] \nn
 &&~\equiv
 S_{d+1}\left[\Gb_{\mu\nu}(x,r;G(x),r_0),\,
  \phib^i(x,r;\phi(x),r_0),\,\Pib^{\mu\nu}(x,r),\,\Pib_i(x,r),\,
  N(x,r),\,\lambda^\mu(x,r)\right] \nn
 &&~= \int d^d x \int_{r_0}dr \,\sqrt{\Gb}\,\,
 \left[\,\Pib^{\mu\nu}\,\dot{\Gb}_{\mu\nu}\,+\,\Pib_i\,\dot{\phib}{}^i
 \,\right].
\end{eqnarray}
Here we have used the Hamiltonian and momentum constraints, 
$\overline{\cH}=\overline{\cP}_\mu=0$. 
One can see that the variation of the action (\ref{eh;off-shell}) is 
given by
\begin{eqnarray}
 \delta S[G(x),\phi(x),r_0]
 &=&-\,\int d^d x\,\sqrt{\Gb}\,\Biggl[\,\left( 
  \Pib^{\mu\nu}(x,r_0)\,\dot{\Gb}_{\mu\nu}(x,r_0)
  +\Pib_i(x,r_0)\,\dot{\phib}{}^i(x,r_0)\right)
 \delta r_0 \nn
 &&~~+\,\Pib^{\mu\nu}(x,r_0)\,\delta\Gb_{\mu\nu}(x,r_0)
     \,+\,\Pib_i(x,r_0)\,\delta\phib^i(x,r_0)\,\Biggr] \nn
 &=&-\,\int d^d x \sqrt{G}\left[\, 
  \Pib^{\mu\nu}(x,r_0)\,\delta G_{\mu\nu}(x)
  +\Pib_i(x,r_0)\,\delta\phi^i(x)
 \,\right],
\end{eqnarray}
since $\delta\Gb_{\mu\nu}(x,r_0)=\delta G_{\mu\nu}(x)
-\dot{\Gb}_{\mu\nu}(x,r_0)\,\delta r_0$, {\em etc}.
It thus follows that the classical conjugate momenta evaluated 
at $r=r_0$ are given by 
\begin{equation}
 \Pi^{\mu\nu}(x)\equiv\Pib^{\mu\nu}(x,r_0)
  ={-1 \over \sqrt{G}}\,{\delta S \over \delta G_{\mu\nu}(x)},
 \qquad
 \Pi_i(x)\equiv\Pib_i(x,r_0)
  ={-1 \over \sqrt{G}}\,{\delta S \over \delta \phi^i(x)}. 
\label{momentum;hj}
\end{equation}
We also see that
\begin{equation}
{\partial \over \partial r_0}S[G_{\mu\nu}(x),\phi^i(x),r_0]=0. 
\end{equation}
Therefore, the on-shell action $S$ is independent of the coordinate 
value of the boundary, $r_0$. 
Substituting (\ref{momentum;hj}) into the Hamiltonian constraint 
(\ref{h;constraint}), 
we thus obtain the flow equation 
of de Boer, Verlinde and Verlinde \cite{dVV}, 
\begin{equation}
\{S,S \}(x)=\sqrt{G(x)}\,\cL_d(x),
\label{hj}
\end{equation}
with
\begin{eqnarray}
 \{S,S \}(x)&\equiv&
 {1 \over \sqrt{G}}\left[ \,
 -\,{1 \over d-1}
  \left( G_{\mu\nu}{\delta S \over \delta G_{\mu\nu}}\right)^2
 +\left( {\delta S \over \delta G_{\mu\nu}}\right)^2
 +{1 \over 2}L^{ij}(\phi)\,{\delta S \over \delta \phi^i}\,
  {\delta S \over \delta \phi^j}
 \,\right], \nonumber \\
 \\
 \cL_d(x)&\equiv&V(\phi)-R+{1 \over 2}\,L_{ij}(\phi)\,G^{\mu\nu}
  \partial_{\mu}\phi^i\partial_{\nu}\phi^j.
\end{eqnarray}
The momentum constraint (\ref{m;constraint})
ensures the invariance of $S$ under a $d$-dimensional diffeomorphism 
along the fixed time slice $r=r_0$: 
\ba
 \int d^d x\,\sqrt{G}\,
  \left[\left(\nabla_\mu\epsilon_\nu+\nabla_\nu\epsilon_\mu
  \right)\frac{\delta S}{\delta G_{\mu\nu}}+\epsilon^\mu\,\partial_\mu
  \phi^i\,\frac{\delta S}{\delta\phi^i}\right]=0,
\ea
with $\epsilon^\mu(x)$ an arbitrary function.

%
\resection{Solution to the flow equation and the Weyl anomaly}

In this section, we discuss a systematic prescription for solving 
the flow equation \eq{hj}.

First we assume that the on-shell action takes the form
\begin{equation}
 S[G(x),\phi(x)]=S_{\rm loc}[G(x),\phi(x)]+\Gamma[G(x),\phi(x)], 
\label{eh;on-shell}
\end{equation}
where $\Sloc[G,\phi]$ is part of $S[G,\phi]$ 
and can be expressed as a sum of local terms:
\begin{eqnarray}
  S_{\rm loc}[G(x),\phi (x)]
  &=&\int d^d x \,\sqrt{G(x)}\,\cLloc(x) \nn
  &=&\int d^d x\,
  \sqrt{G(x)}\sum_{w=0,2,4,\cdots}\bigl[\cLloc(x)\bigr]_w.
\end{eqnarray}
Here we have arranged the sum over local terms 
according to the weight $w$ that is defined additively 
from the following rule\footnote{
A scaling argument of this kind is often used in supersymmetric theories
to restrict the form of low energy effective actions 
(see e.g.\ Ref.\ \cite{GSW}).
}:
\begin{center}
\begin{tabular}{c|c}
       & weight \\ \hline
$G_{\mu\nu}(x), \,\phi^i(x), \,\Gamma[G,\phi]$ & $0$ \\ \hline
$\partial_{\mu}$ & $1$ \\ \hline
$R, \,R_{\mu\nu},\, \partial_\mu\phi^i\partial_\nu\phi^j,\,
\cdots$ & $2$ \\ \hline
$\delta \Gamma / \delta G_{\mu\nu}(x),\, 
\delta \Gamma / \delta \phi^i(x)$ & $d$ 
\end{tabular}
\end{center}
The last line is a natural consequence of the relation
$w\!\left(\Gamma[G,\phi]\right)=0$, 
since $\delta\Gamma=\int d^d x \linebreak \left(\delta\phi(x)\times
\delta\Gamma/\delta\phi(x)+\cdots\right)$.
Then, substituting the above equation into the flow equation (\ref{hj}) 
and comparing terms of the same weight, 
we obtain a sequence of equations that relate the off-shell bulk
action (\ref{eh;off-shell}) to the on-shell boundary 
action (\ref{eh;on-shell}). 
They take the following form:
\ba
 \sqrt{G}\,\cL_d&=&\Bigl[\left\{\Sloc,\,\Sloc\right\}\Bigr]_0
  +\Bigl[\left\{\Sloc,\,\Sloc\right\}\Bigr]_2\label{eqA}\,,\\
 0&=&\Bigl[\left\{\Sloc,\,\Sloc\right\}\Bigr]_w
  \quad(w=4,6,\cdots,d-2), \label{eqB} \\
 0&=&2\Bigl[\left\{\Sloc,\,\Gamma\right\}\Bigr]_d
  +\Bigl[\left\{\Sloc,\,\Sloc\right\}\Bigr]_d \label{eqC}\,,\\
 0&=&2\Bigl[\left\{\Sloc,\,\Gamma\right\}\Bigr]_w
  +\Bigl[\left\{\Sloc,\,\Sloc\right\}\Bigr]_w 
  \quad(w=d+2,\cdots,2d-2)\label{eqD}, \\
 0&=&\Bigl[\left\{\Gamma,\,\Gamma\right\}\Bigr]_{2d}
  +2\Bigl[\left\{\Sloc,\,\Gamma\right\}\Bigr]_{2d}
  +\Bigl[\left\{\Sloc,\,\Sloc\right\}\Bigr]_{2d}\,,\\
 0&=&2\Bigl[\left\{\Sloc,\,\Gamma\right\}\Bigr]_w
  +\Bigl[\left\{\Sloc,\,\Sloc\right\}\Bigr]_w 
  \quad(w=2d+2,\cdots). \label{eqE}\,
\ea
Equations \ \eq{eqA} and \eq{eqB} determine 
$\left[\cLloc\right]_w~(w=0,2,\cdots,d-2)$, 
which together with Eq.\ \eq{eqC} in turn determine the non-local 
functional $\Gamma$. 
Although $\left[\cLloc\right]_d$ could enter the expression, 
this would not give a physically relevant effect, 
as we see below.

By parametrizing $[\cLloc]_0$ and $[\cLloc]_2$ as 
\begin{eqnarray}
  \left[\cLloc\right]_0&=& W(\phi), \\
  \left[\cLloc\right]_2&=& -\Phi(\phi)\,R+{1 \over 2}\,M_{ij}(\phi)\,
   G^{\mu\nu}\,\partial_{\mu}\phi^i\,\partial_{\nu}\phi^j,
\end{eqnarray}
one can easily solve \eq{eqA} to obtain\footnote{
The expression for $d=4$ can be found in Ref.\ \cite{dVV}.
}
\begin{eqnarray}
 V(\phi)&=&-{d \over 4(d-1)}\,W(\phi)^2
  +{1 \over 2}L^{ij}(\phi)\,\partial_i W(\phi)\,\partial_j W(\phi)\,, 
  \label{hj;potential} \\
 -1&=&{d-2 \over 2(d-1)}\,W(\phi)\,\Phi(\phi)
  -L^{ij}(\phi)\,\partial_i W(\phi)\,\partial_j \Phi(\phi)\,, 
  \label{hj;kineG} \\
 {1 \over 2}\,L_{ij}(\phi)&=&-{d-2 \over 4(d-1)}\,W(\phi)\,M_{ij}(\phi)
  -L^{kl}(\phi)\,\partial_k W(\phi)\,\Gamma^{(M)}_{l;ij}(\phi)\,,
 \label{hj;kinesca}\\
 0&=&W(\phi)\,\nabla^2\,\Phi(\phi)+L^{ij}(\phi)\,\partial_i W(\phi)\,
  M_{jk}(\phi)\,\nabla^2\phi^k\,.
\end{eqnarray}
Here $\partial_i=\partial/\partial\phi^i$, and 
$\Gamma^{(M)k}_{ij}(\phi)\equiv 
M^{kl}(\phi)\,\Gamma^{(M)}_{l;ij}(\phi)$ is the Christoffel symbol 
constructed from $M_{ij}(\phi)$. 
For pure gravity ($L_{ij}=0, M_{ij}=0$), for example,  
setting $V=2\Lambda=-d(d-1)/l^2$, 
we find\footnote{
The sign of $W$ is chosen to be in the branch 
where the limit $\phi\rightarrow 0$ can be taken smoothly
with $L_{ij}(\phi)$ and $M_{ij}(\phi)$ positive definite.
} 
\ba
  W=-\,\frac{2\,(d-1)}{l},\quad \Phi=\frac{l}{d-2}.
\ea 
Here $\Lambda$ is the bulk cosmological constant, 
and when the metric is asymptotically AdS, 
the parameter $l$ is identified with the radius of 
the asymptotic AdS${}_{d+1}$.

To solve Eq.\ \eq{eqB}, we need to introduce local terms of 
higher weight ($w\geq4$). 
For example, for the pure gravity case, 
we need a local term $[\cLloc]_4$ of the form 
\begin{eqnarray}
 [\cLloc]_4=XR^2+YR_{\mu\nu}R^{\mu\nu}
 +ZR_{\mu\nu\lambda\sigma}R^{\mu\nu\lambda\sigma},
\label{l4}
\end{eqnarray}
with $X,Y$ and $Z$ being some constants to be determined. 
By using this, we find that 
\begin{eqnarray}
  &&\frac{1}{\sqrt{G}}\Bigl[
  \left\{ S_{\rm loc}, S_{\rm loc} \right\}\Bigr]_4 \nn
  &&~=-{W \over 2(d-1)} 
   \left( (d-4)X-{d\,l^3 \over 4(d-1)(d-2)^2}\right) R^2\nn
  &&~~~~~ -{W \over 2(d-1)} \left( (d-4)Y+{l^3 \over (d-2)^2}\right)
   R_{\mu\nu}R^{\mu\nu}~-{ d-4 \over 2(d-1)}\,W Z\,
   R_{\mu\nu\lambda\sigma}R^{\mu\nu\lambda\sigma}\ \nn
  &&~~~~~ +\left( 2X+{d \over 2(d-1)}Y+{2 \over d-1}Z\right)\nabla^2R. 
\label{slsl;w4}
\end{eqnarray}
Thus for $d\geq6$, requiring 
$\Bigl[\{ S_{\rm loc}, S_{\rm loc} \}\Bigr]_4=0$, 
we have 
\begin{equation} 
 X={d\,l^3 \over 4(d-1)(d-2)^2(d-4)},\quad 
 Y=-{l^3 \over (d-2)^2(d-4)},\quad
 Z=0. \label{d=6}
\end{equation}
Note that the coefficient of $\nabla^2R$ vanishes. 
From Eq.\ (\ref{d=6}), $\Bigl[\{\Sloc,\Sloc\}\Bigr]_6$ 
can be calculated easily to be
\begin{eqnarray}
 &&\frac{1}{\sqrt{G}}\Bigl[
  \left\{ S_{\rm loc}, S_{\rm loc} \right\}\Bigr]_6 \nn
 &&~~=
  \Phi\left[\,\left(-4X+\frac{d+2}{2(d-1)}Y\right)
    R R_{\mu\nu} R^{\mu\nu}
  +{d+2\over 2(d-1)}XR^3
  -4\,YR^{\mu\lambda} R^{\nu\sigma} R_{\mu\nu\lambda\sigma}\right.\nn
 &&~~~~\left.+(4X+2Y)R^{\mu\nu}\nabla_{\mu}\nabla_{\nu}R
  -2YR^{\mu\nu}\nabla^2R_{\mu\nu}
  +\left(2(d-3)X+\frac{d-2}{2}Y\right)R\,\nabla^2R
 \,\right]\nn
 &&~~~~+({\rm contributions~from}~[\cLloc]_6)\nn
 &&~~=
  l^4\left[\,-\,\frac{3d+2}{2(d-1)(d-2)^3(d-4)}
   \,RR_{\mu\nu}\,R^{\mu\nu}
  +{d(d+2)\over 8(d-1)^2(d-2)^3(d-4)}\,R^3 \right.\nn
 &&~~~~+\,{4\over (d-2)^3(d-4)}\,R^{\mu\lambda}\,R^{\nu\sigma}\,
    R_{\mu\nu\lambda\sigma}
  -\,{1\over (d-1)(d-2)^2(d-4)}\,
   R^{\mu\nu}\,\nabla_{\mu}\nabla_{\nu}R\nn
 &&~~~~\left.  +\,{2\over (d-2)^3(d-4)}\,R^{\mu\nu}\,\nabla^2R_{\mu\nu}
  -\,{1\over (d-1)(d-2)^3(d-4)}\,R\,\nabla^2R
 \,\right]\nn
 &&~~~~+\,({\rm contributions~from}~[\cLloc]_6).
\label{slsl;w6}
\end{eqnarray}

On the other hand, from Eq.\ \eq{eqC} in the flow equation 
with weight $d$, we find
\ba
  -2 [\gamma]_0 \,G_{\mu\nu} {\delta\Gamma \over \delta G_{\mu\nu}}
 +[\gamma B_{\mu\nu}]_0\, {\delta\Gamma \over \delta G_{\mu\nu}}
 +[\gamma B^i]_0\, {\delta \Gamma \over \delta \phi^i} 
 =-\Bigl[\{ S_{\rm loc},S_{\rm loc} \}\Bigr]_d,
\ea
where
\begin{eqnarray}
 \gamma &\equiv&
  \frac{1}{d(d-1)}\,\frac{1}{\sqrt{G}}\,G_{\mu\nu}\,
  \frac{\delta \Sloc}{\delta G_{\mu\nu}}\nn
  &=& {1 \over d(d-1)}\left({d \over 2}\,
 W-{d-2 \over 2}\,R\,\Phi -(d-1)\,\nabla^2\Phi
 +{d-2 \over 4}\,M_{ij}\,G^{\mu\nu}\,
  \partial_{\mu}\phi^i\,\partial_{\nu}\phi^j \right) \,,\nn
 && \\
 \gamma B_{\mu\nu} &\equiv&
  \frac{2}{\sqrt{G}}\left(G_{\mu\lambda}G_{\nu\sigma}\,
     -\frac{1}{d}\,G_{\mu\nu}G_{\lambda\sigma}\right)\,
  \frac{\delta \Sloc}{\delta G_{\lambda\sigma}}
  \qquad\Bigl(G^{\mu\nu}B_{\mu\nu}=0\Bigr)\nn
  &=&  2\,\Phi\, R_{\mu\nu}-{2 \over d}\,G_{\mu\nu}\,\Phi R
  +2\,\nabla_{\mu}\nabla_{\nu}\Phi-{2 \over d}\,G_{\mu\nu}\,\nabla^2\Phi 
  \nn
 &&~~~
 +{1\over d}\,G_{\mu\nu}\,M_{ij}\,
  \partial_{\sigma}\phi^i\,\partial^{\sigma}\phi^j
  -M_{ij}\,\partial_{\mu}\phi^i\,\partial_{\nu}\phi^j\,, \\ 
 \gamma B^i  &\equiv& 
  \frac{1}{\sqrt{G}}\,L^{ij}(\phi)\,\frac{\delta\Sloc}{\delta\phi^j}\nn
  &=& L^{ij}\left( \partial_jW
  -\partial_j\Phi\,R-M_{ij}\,\nabla^2\phi^k
  -\Gamma^{(M)}_{j;kl}\,\partial_{\mu}\phi^k\,
  \partial^{\mu}\phi^l \right)\,. 
\end{eqnarray}
Since $[B_{\mu\nu}]_0=0$, we have
\ba
  -2 G_{\mu\nu} {\delta\Gamma \over \delta G_{\mu\nu}}
  +\beta^i\, {\delta \Gamma \over \delta \phi^i} 
 ={-1\over [\gamma]_0}\,\Bigl[\{ S_{\rm loc},S_{\rm loc} \}\Bigr]_d,
\label{cs;functional}
\ea
with
\begin{equation}
 [\gamma]_0={W(\phi) \over 2(d-1)},
 \qquad \beta^i\equiv [B^i]_0
  ={2(d-1) \over W(\phi)}L^{ij}(\phi)\,\partial_j W(\phi). 
\label{cs;beta}
\end{equation}
As we see below, $\beta^i$ can be identified with the RG beta function, 
so that the right-hand side of (\ref{cs;functional}) (divided by
$\sqrt{G}$) expresses the Weyl anomaly $\cW_d$ of the $d$-dimensional
boundary field theory: 
\ba
 {1\over [\gamma]_0}\,
  \Bigl[\{ S_{\rm loc},S_{\rm loc} \}\Bigr]_d
  =2\sqrt{G}\left(\cW_d + \nabla_\mu \cJ_d^\mu\right), 
\ea
or
\ba
 \cW_d + \nabla_\mu \cJ_d^\mu=
  \frac{d-1}{W(\phi)\sqrt{G}}\,\Bigl[\left\{\Sloc,\Sloc\right\}\Bigr]_d,
\ea
where the total derivative term $\nabla_\mu\cJ_d^\mu$ represents 
the contribution from $[\cLloc]_d$.
The fact that the effect of $[\cLloc]_d$ can always be 
put into the form of a total derivative  
can be seen directly for pure gravity in five dimensions. 
In fact, setting $d=4$ in Eq.\ \eq{slsl;w4}, 
the dependence on $X,Y$ and $Z$ (coming from 
$[\cLloc]_4$) totally disappears, 
except for the last total derivative term. 
This can be generally understood by observing 
that possible contributions from $[\cLloc]_d$ always vanish 
for constant dilatations.

To illustrate how the above prescription works, 
we consider two simple cases.

\noindent\underline{\bf 5D dilatonic gravity}:

We normalize the Lagrangian with a single scalar field
as follows:
\ba
  \cL_4=-\frac{12}{l^2}-R+\frac{1}{2}
  \,G^{\mu\nu}\,\partial_\mu\phi\,\partial_\nu\phi.
\ea
Then, assuming that all the functions $W(\phi), M(\phi)$ and 
$\Phi(\phi)$ are constant in $\phi$,
we can solve Eqs.\ \eq{hj;potential}--\eq{hj;kinesca} with  
$V=-d(d-1)/l^2=-12/l^2$ and $L=1$, 
and obtain 
\ba
  W=-\frac{6}{l},\quad \Phi=\frac{l}{2},\quad M=\frac{l}{2};
\ea
that is, 
\begin{equation}
 S_{{\rm loc}}[G,\phi]=\int d^4x\sqrt{G}\left( -\frac{6}{l}-{l\over 2}R
 +{l\over 2}G^{\mu\nu}\partial_{\mu}\phi\,\partial_{\nu}\phi
\right).
\label{sloc4;dvv}
\end{equation}
We can calculate $\Bigl[\{\Sloc,\Sloc\}\Bigr]_4$ easily 
and find 
\begin{eqnarray}
 \cW_4&=&-\frac{l}{2\sqrt{G}}\,\Bigl[\{\Sloc,\Sloc\}\Bigr]_4\nn
 &=&l^3\left({1\over 24}R^2-{1\over 8}R_{\mu\nu}R^{\mu\nu}
  -{1\over 24}R\,G^{\mu\nu}\,\partial_{\mu}\phi\,
  \partial_{\nu}\phi\right. \nn
 &&\left.+{1\over 8}R^{\mu\nu}\partial_{\mu}\phi\,
  \partial_{\nu}\phi 
  -{1\over 48}\left(G^{\mu\nu}\partial_{\mu}\phi\,
   \partial_{\nu}\phi\right)^2
 -{1\over 16}\left(\nabla^2\phi\right)^2\right) .
 \label{weyl4}
\end{eqnarray}
This is in exact agreement with the result in Ref.\ \cite{dilaton}.

\noindent\underline{\bf 7D pure gravity}:

By using the value in Eq.\ \eq{d=6} with $d=6$, the local part of 
weight up to four is given by
\begin{equation}
S_{\rm loc}[G]=\int d^6x \sqrt{G}\left( -\frac{10}{l}-{l\over 4}R 
+{3l^3\over 320}R^2-{l^3\over 32}R_{\mu\nu}R^{\mu\nu}\right). 
\label{sloc6;dvv}
\end{equation}
{}From the flow equation of weight $w=6$, 
we thus find
\begin{eqnarray}
 \cW_6&=&-\frac{l}{2\sqrt{G}}\Bigl[\left\{\Sloc,\Sloc\right\}\Bigr]_6 \nn
 &=&l^5\left({1\over 128}\,RR_{\mu\nu}R^{\mu\nu}-{3\over 3200}\,R^3
  -{1\over 64}\,R^{\mu\lambda}R^{\nu\sigma}R_{\mu\nu\lambda\sigma}\right.\nn
 &&~~\left.+\,{1\over 320}\,R^{\mu\nu}\nabla_{\mu}\nabla_{\nu}R
  -{1\over 128}\,R^{\mu\nu}\nabla^2R_{\mu\nu}+{1\over 1280}\,
   R\,\nabla^2R\right),
 \label{weyl6}
\end{eqnarray}
which is in perfect agreement with the six-dimensional Weyl anomaly 
given in Ref.\ \cite{HS;weyl}.

We conclude this section by showing that one can generalize 
to arbitrary dimension the argument in Ref.\ \cite{dVV} 
that the scaling dimension can be calculated directly from the
flow equation. 
{}First, we assume that the scalars are normalized as
$L_{ij}(\phi)=\delta_{ij}$ 
and that the bulk scalar potential 
$V(\phi)$ has the expansion
\begin{equation}
 V(\phi)=2\Lambda+{1\over 2}\sum_i m_i^2\,\phi_i^2
 +\sum_{ijk}g_{ijk}\,\phi_i\phi_j\phi_k +\cdots ,
\end{equation}
with $\Lambda=-d(d-1)/2l^2$.
Then it follows from (\ref{hj;potential}) 
that $W$ takes the form 
\begin{equation}
 W=-\frac{2(d-1)}{l}+{1\over 2}\sum_i \lambda_i\,\phi_i^2
 +\sum_{ijk}\lambda_{ijk}\,\phi_i\phi_j\phi_k+\cdots,
\end{equation}
with 
\begin{eqnarray}
 &&l\lambda_i={1\over 2}\left( -d + \sqrt{ d^2+4\,m_i^2\,l^2}\right), \\
 &&g_{ijk}= \left(\frac{d}{l}
 +\lambda_i+\lambda_j+\lambda_k\right)\lambda_{ijk}.
\end{eqnarray}
{}Furthermore, if we perturb the system finitely 
by fixing the sources $\phi^i(x)$ to be constant 
and fixing the form of $G_{\mu\nu}(x)$ as $\delta_{\mu\nu}/a^2$ 
with some constant $a$, 
then the functions $\beta^i$ can be regarded 
as the beta functions with $a$ being the cutoff length, 
as shown in Ref.\ \cite{dVV} (see also Appendix C). 
They can be evaluated easily and are found to be 
\begin{equation}
 \beta^i=-\sum_i l\lambda_i\,\phi_i
   -3\sum_{jk}\lambda_{ijk}\,\phi_j\phi_k+\cdots. 
\end{equation}
Thus, equating the coefficient of the first term 
with $d-\Delta_i$, where $\Delta_i$ is the scaling dimension of the operator 
coupled to $\phi_i$, 
we thus obtain 
\begin{equation}
 \Delta_i=d+l\lambda_i={1\over 2}
  \left( d + \sqrt{ d^2+4\,m_i^2\,l^2 }\right). 
\end{equation}
This exactly reproduces the result given in Ref.\ \cite{MGKPW}.

%
\resection{Continuum limit}

In this section, we describe a direct prescription for taking 
continuum limits of boundary field theories which is such that 
counterterms can be extracted easily.\footnote{
{}For earlier work on counterterms, see e.g.\ Ref.\ \cite{BK}.
}

Let $\Gb_{\mu\nu}(x,r;G(x),r_0)$ and $\phib^i(x,r;\phi(x),r_0)$
be the classical trajectory of $G_{\mu\nu}(x,r)$ 
and $\phi^i(x,r)$ with the boundary condition 
\begin{equation}
 \Gb_{\mu\nu}(x,r\!=\!r_0)=G_{\mu\nu}(x),\qquad
 \phib^i(x,r\!=\!r_0)=\phi^i(x).
\end{equation}
Recall that the on-shell action is given as a functional of 
the boundary values $G_{\mu\nu}(x)$ and $\phi^i(x)$, 
obtained by substituting these classical solutions into the bulk action:
\begin{equation}
 S[G_{\mu\nu}(x),\phi^i(x)]=\int d^d x \int_{r_0}dr\,\sqrt{\Gb}\,
 {\cal L}_{d+1}\left[\Gb(x,r;G,r_0),\phib(x,r;\phi,r_0)\right]. 
\end{equation}
Also, recall that the fields $G_{\mu\nu}(x)$ and $\phi^i(x)$ are 
considered as the bare sources at the cutoff scale corresponding to 
the flow parameter $r_0$. 
Although the on-shell action is actually independent of  
$r_0$ due to the Hamilton-Jacobi constraint, 
we still need to tune the fields $G_{\mu\nu}(x)$ and $\phi^i(x)$ 
as functions of $r_0$ so that 
the low energy physics is fixed and described in terms of finite 
renormalized couplings.

In the holographic RG \cite{dVV}, such renormalization can be 
easily carried out 
by tuning the bare sources back along the classical trajectory 
in the bulk (see Fig.\ 1). 
\begin{figure}
\begin{center}
\leavevmode
\includegraphics{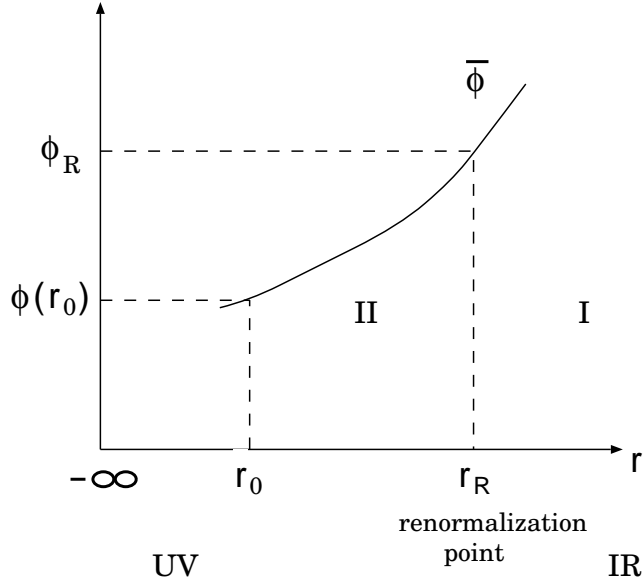}
\end{center}
\caption{\footnotesize{The evolution of the classical solutions 
$\phib^i$ along the radial direction. The region I is defined 
by $r\ge r_R$, and the region II is defined by $r_0\le r < r_R$. }}
\label{renormalize}
\end{figure}
That is, if we would like to fix the couplings at the
``renormalization point'' $r=r_R$ 
to be $G_R(x)$ and $\phi_R(x)$ and to require that physics 
does not change as the cutoff moves, 
we only need to take the bare sources to be 
\begin{equation}
 G_{\mu\nu}(x,r_0)=\Gb_{\mu\nu}(x,r_0;G_R,r_R),\qquad
 \phi^i(x,r_0)=\phib^i(x,r_0;\phi_R,r_R). 
\label{bare}
\end{equation}

The on-shell action with these running bare sources can be 
easily evaluated by using Eq.\ (\ref{bare}):
\begin{eqnarray}
S[G_{\mu\nu}(x,r_0),\,\phi^i(x,r_0)] \!\!\!&=&\!\!\! \int d^dx\int_{r_0}dr 
\,\sqrt{\Gb}\,{\cal L}_{d+1}
\left[\Gb(x,r;G_R,r_R),\phib(x,r;\phi_R,r_R)\right] 
\nonumber \\
\!\!\!&=&\!\!\!\int d^dx\left( \int_{r_R}dr+\int_{r_0}^{r_R}dr \right)
\sqrt{\Gb}\,{\cal L}_{d+1} \nonumber \\
\!\!\!&=&\!\!\!\ S_R[G_R(x),\phi_R(x)]
+S_{\rm CT}\left[G_R(x),\phi_R(x);r_0,r_R\right].
\label{counter}
\end{eqnarray}
Here $S_R$ is given by integrating 
$\sqrt{\Gb}{\cal L}_{d+1}$ over the region I in Fig.\ 
\ref{renormalize}, and it obeys the Hamiltonian constraint, 
which ensures that $S_R$ does not depend on $r_R$. 
On the other hand, $S_{\rm CT}$ is given by integrating 
$\sqrt{\Gb}{\cal L}_{d+1}$ over the region II. 
{ }It also obeys the Hamiltonian constraint 
and thus does not depend on the coordinates of the boundaries 
of integration, $r_R$ and $r_0$, explicitly. 
However, in this case, their dependence implicitly enters  
$S_{\rm CT}$ through the condition that the boundary values at $r=r_0$
are on the classical trajectory through the renormalization point:
\ba
 S_{{\rm CT}}&=&S\left[G_R(x),\phi_R(x);G(x,r_0),\phi(x,r_0)\right] \nn
 &=&S\left[G_R(x),\phi_R(x);\Gb(x,r_0;G_R,r_R),
 \phib(x,r_0;\phi_R,r_R)\right].
\ea
It is thus natural to interpret $S_{\rm CT}[G_R,\phi_R;r_0,r_R]$ as 
the counterterm, and the nonlocal part of $S_R[G_R,\phi_R]$ gives
the renormalized generating functional of the boundary field theory, 
$\Gamma_R[G_R,\phi_R]$, 
written in terms of the renormalized sources.

Since, as pointed out above, $S_R[G_R,\phi_R]$ also satisfies the Hamiltonian 
constraint, it will yield the same form of the flow equation, 
with all the bare fields replaced by the renormalized fields. 
This suggests\footnote{
We thank H.\ Sonoda for discussions on this point.
}
that the holographic RG exactly describes 
the so-called renormalized trajectory \cite{wilson-kogut},   
which is a submanifold in the parameter space, 
consisting of the flows driven by relevant perturbations 
from an RG fixed point at $r_0=-\infty$.

%
\resection{Relation to the analysis by Henningson and Skenderis}

In this section, we comment on the relation between the analysis given 
above and that of Henningson and Skenderis \cite{HS;weyl}, 
which is briefly reviewed in Appendix D. 
In particular, we show that $\Sloc$, the local part of 
the on-shell action, can also be calculated 
solely from their analysis. 
In the following discussion, we exclusively consider 
the pure gravity case. 
Extension to the case in which matter fields exist 
should be straightforward.

{}First, we recall that in our analysis, the bare coupling $G(x)$ 
at $r=r_0$ is tuned in such a way that it is on the classical trajectory 
that passes through a fixed value $G_R$ at some renormalization 
point, $r=r_R$ (see Eq.\ \eq{bare}): 
\ba
 G(x) \rightarrow G(x,r_0)=\Gb(x,r_0;G_R,r_R).
\ea
The value $G_R$ is regarded as the renormalized coupling at $r=r_R$. 
On the other hand, it is also possible to choose 
as the renormalized coupling the coefficient of  
the asymptotic form of the classical solution, 
as is done in Ref.\ \cite{HS;weyl}. 
That is, by expanding the classical solution 
in the limit $r\rightarrow -\infty$,
\ba
 \Gb_{\mu\nu}(x,r)=e^{-2r}\left(g^{(0)}_{\mu\nu}(x)
  +e^{2r}g^{(2)}_{\mu\nu}(x)+\cdots\right),
\ea
one can interpret $g_{(0)}$ as the renormalized coupling. 
Here $g_{(2)},g_{(4)}, \cdots$ are obtained as local functions
constructed from $g_{(0)}$ in such a way that $\Gb(x,r)$ satisfies 
the equation of motion. 
Some of them are given explicitly in Appendix D.  
The two renormalized couplings, $G_R(x)$ and $g_{(0)}(x)$, 
are related through the simple relation 
\ba
  G(x,r_0)&=&\Gb(x,r_0;G_R,r_R)\nn
  &=&e^{-2r_0}\left(g_{(0)}(x)+e^{2r_0}g_{(2)}\!\left[g_{(0)}(x)\right]
  +e^{4r_0}g_{(4)}\!\left[g_{(0)}(x)\right]+\cdots\right).
\ea

Now we show that once the counterterm is known 
within the scheme of Henningson and Skenderis, 
we can directly calculate the local part of the on-shell action, 
$\Sloc[G]$.
To show this, we first introduce the new coordinate $\rho\equiv e^{2r}$ 
and set $\epsilon\equiv e^{2r_0}$. 
The classical solution is thus expanded around $\rho=0$ as 
(see Appendix D)\footnote{
In the following discussion, 
we write $\Gb(x,r(\rho))~(r(\rho)=(1/2)\log\rho)$ 
simply as $\Gb(x,\rho)$.
}
\ba
 \Gb(x,\rho)\!\!\!&=&\!\!\!\frac{1}{\rho}\,\Biggl[g_{(0)}(x)
  +\rho\,g_{(2)}\!\left[g_{(0)}(x)\right]+\cdots 
  +\rho^{d/2}\left(g_{(d)}\!\left[g_{(0)}(x)\right]
  +\log\rho\,h_{(d)}\!\left[g_{(0)}(x)\right]
  \right)+\cdots\Biggr].\nn
 &&
\ea
We then require that this classical solution passes through 
the point $\gh(x)/\epsilon$ at $\rho=\epsilon$ (see Fig.\ 2) 
with $\gh(x)$ some fixed function:
\ba
 \frac{1}{\epsilon}\,\gh(x)\equiv\Gb(x,\epsilon) 
 =\frac{1}{\epsilon}\,\left(g_{(0)}(x)
  +\epsilon\,g_{(2)}\!\left[g_{(0)}(x)\right]+\cdots\right).
\ea
\begin{figure}
\begin{center}
\leavevmode
\includegraphics{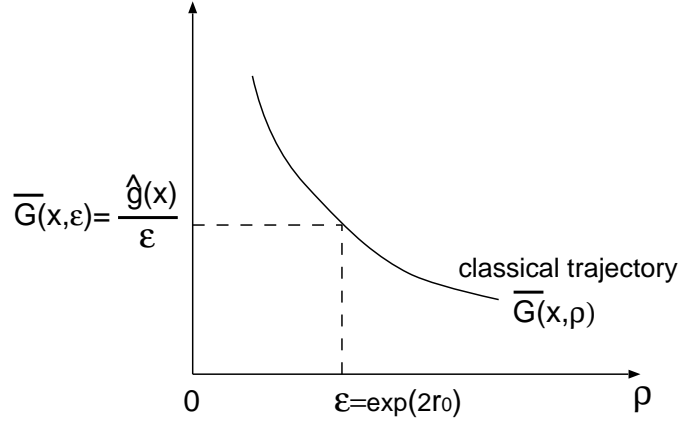}
\end{center}
\caption{\footnotesize{The classical solution $\Gb(x,\rho)$ 
with $\rho=\exp(2r)$ is chosen such that it passes through 
the point $\gh(x)/\epsilon$ at $\rho=\epsilon$ ({\em i.e.}, 
$r=r_0$).}}
\end{figure}
This can be solved recursively as 
\ba
  g_{(0)}\!\left[\gh(x),\epsilon\right]
  =\gh(x)+\epsilon\,b_{(2)}\!\left[\gh(x)\right]
  +\cdots
  +\epsilon^{d/2}\,
  \left(b_{(d)}\!\left[\gh(x)\right]
  +\log\epsilon\,c_{(d)}\!\left[\gh(x)\right]
  \right)+\cdots.
\ea
Since $G=\gh/\epsilon$ is the boundary value of the classical solution 
at $\rho=\epsilon$ ({\em i.e.}, $r=r_0$), 
we have
\begin{equation}
S\left[\gh(x)/\epsilon \right]
=S_{d+1}\left[\Gb\left(x,r;\,\gh/\epsilon,r_0\right)\right].
\end{equation}
The right-hand side is identical to the on-shell action 
in the scheme of Henningson and Skenderis given in Appendix D, 
with $g_{(0)}=g_{(0)}[\gh,\epsilon ]$. 
We thus have
\begin{equation}
 S[\gh/\epsilon]
  =S^{\rm HS}\left[ g_{(0)}[\gh,\epsilon],\epsilon \right].
\label{hsdvv}
\end{equation}
We can then extract the terms that diverge in the limit 
$\epsilon\rightarrow 0$ as follows.
We first note that $S[\gh/\epsilon]$ can be written as 
\begin{equation}
 S[\gh/\epsilon]=\Sloc[\gh/\epsilon]+\Gamma[\gh/\epsilon]. 
\end{equation}
Here $\Sloc[\gh/\epsilon]$ is a meromorphic function of 
$\epsilon$ and has the following Laurent expansion:
\begin{equation}
S_{\rm loc}[\gh/\epsilon]=\sum_{w=0,2,4,\cdots}\epsilon^{(w-d)/2} 
\int d^dx\sqrt{\gh}\Bigl[{\cal L}_{\rm loc}[\gh]\Bigr]_w. 
\end{equation}
$\Gamma[\gh/\epsilon]$, on the other hand, may lead to a 
logarithmically divergent term. 
We thus obtain the following equation for the divergent terms:
\begin{equation}
\sum_{w=0}^{d-2}\epsilon^{(w-d)/2} 
\int d^dx\,\sqrt{\gh}\,\Bigl[{\cal L}_{\rm loc}[\gh]\Bigr]_w 
  -\log\epsilon\,\int d^d x\,\sqrt{\gh}\,\,\widehat{\cW}_d
  \,=\,S^{\rm HS}_{\rm div}\left[
   g_{(0)}[\gh,\epsilon],\,\epsilon \right].
 \label{fmshs}
\end{equation}
The quantity $S^{\rm HS}_{\rm div}\left[g_{(0)}[\gh,\epsilon],\,\epsilon
\right]$, the divergent part of $S^{\rm HS}$, 
is calculated in Ref.\ \cite{HS;weyl} (see also Appendix D).  
By considering the structure, one can easily understand that 
$\widehat{\cW}_d$ should be the Weyl anomaly 
written in terms of $\gh$. 
Equation \eq{fmshs} shows that the relevant part of $\Sloc$ can be 
calculated from the divergent term of $S^{\rm HS}$ 
by comparing terms of the same order in $\epsilon$.

We now give sample calculations for $d=4$ and $d=6$. 

\noindent\underline{\bf $d=4$}:

Straightforward calculation gives the coefficients 
$b_{(2)},\cdots$ as
\begin{eqnarray}
\begin{array}{cclcl}
 b_{(2)}&=&-\gh_{(2)} & &\left(\gh_{(2)}\equiv g_{(2)}[\gh]\right), \\
 b_{(4)}&=&-\Dh-\gh_{(4)} & & \left(\gh_{(4)}\equiv g_{(4)}[\gh]\right), \\
 c_{(4)}&=&-\hh_{(4)} & & \left(\hh_{(4)}\equiv h_{(4)}[\gh]\right), 
\end{array}
\end{eqnarray}
where $\widehat{D}_{\mu\nu}$ is the covariant tensor given by 
\begin{eqnarray}
\Dh_{\mu\nu}&=&{1\over 4}\left[\, \widehat{\nabla}^{\sigma}\left(
\widehat{\nabla}_{\mu}\gh^{(2)}_{\nu\sigma}
+\widehat{\nabla}_{\nu}\gh^{(2)}_{\mu\sigma}\right)
-\widehat{\nabla}^2\gh^{(2)}_{\mu\nu}
-\widehat{\nabla}_{\mu}\widehat{\nabla}_{\nu}
\tr (\gh^{-1}\gh_{(2)})\,\right] \nonumber \\
&&~-{1\over 12}\left[\, 
-\gh^{(2)}_{\lambda\sigma}\widehat{R}^{\lambda\sigma}
+\widehat{\nabla}^{\lambda}\widehat{\nabla}^{\sigma}
\gh^{(2)}_{\lambda\sigma}
-\widehat{\nabla}^2\tr (\gh^{-1}\gh_{(2)})\, \right]\gh_{\mu\nu}
-{1\over 12}\,\widehat{R}\,\gh^{(2)}_{\mu\nu}. \nonumber \\
\end{eqnarray}
Substituting these values into Eq.\ \eq{sloc4;hs},
we obtain
\begin{eqnarray}
 S_{\rm div}^{\rm HS}\left[g_{(0)}[\gh(x),\epsilon],\epsilon\right] 
 =\int d^4x\sqrt{\gh}
 \left( -{6\over\epsilon^2}-{1\over 2\epsilon}\widehat{R}
 -\log\epsilon\,\,\widehat{\cW}_4
 \right).
\label{sloc4;fms}
\end{eqnarray}
This actually gives Eqs.\ \eq{sloc4;dvv} and \eq{weyl4} 
with $\phi=0$ and $l=1$.

\noindent\underline{\bf $d=6$}:

The coefficients are calculated to be 
\begin{eqnarray}
 b^{(2)}_{\mu\nu}&=&{1\over 4}\left( \widehat{R}_{\mu\nu}
  -{1\over 10}\widehat{R}\gh_{\mu\nu}\right), \nonumber \\
 b^{(4)}_{\mu\nu}&=&{1\over 16}\widehat{R}^{\lambda\sigma}
  \widehat{R}_{\mu\lambda\nu\sigma}
  -{1\over 32}\widehat{R}_{\mu}^{\lambda}\widehat{R}_{\lambda\nu}
  -{1\over 80}\widehat{R}\widehat{R}_{\mu\nu}-
  {9\over 640}\gh_{\mu\nu}\widehat{R}^{\lambda\sigma}
  \widehat{R}_{\lambda\sigma}
  +{9\over 4^4 5^2 }\widehat{R}^2\,\gh_{\mu\nu}, \nonumber \\
\end{eqnarray}
which lead to 
\begin{eqnarray}
 S^{\rm HS}_{\rm div}\left[g_{(0)}[\gh(x),\epsilon],\epsilon\right]
 \!\!\!&=&\!\!\!\int d^6x\sqrt{\gh}
 \left( -{10 \over \epsilon^3}-{1\over 4\epsilon^2}\widehat{R}
 +{3\over 320\epsilon}\widehat{R}^2-{1\over 32\epsilon}
 \widehat{R}_{\mu\nu}\widehat{R}^{\mu\nu}
 -\log\epsilon\,\,\widehat{\cW}_6\right).\nn
 &&
\end{eqnarray}
This reproduces Eqs.\ \eq{sloc6;dvv} and \eq{weyl6} with $l=1$.

%
\resection{Conclusion}

In this note, we have discussed several aspects of the holographic RG 
that are related to the Weyl anomaly. 
We found that the Hamilton-Jacobi constraint is quite useful 
in exploring the holographic RG, 
especially to calculate the Weyl anomaly 
and to understand the structure of divergent parts. 
We also discussed continuum limits of the boundary theories 
in the context of the holographic RG. 
In particular we demonstrated that counterterms can be 
extracted systematically 
if we use a special renormalization, 
where the bare and the renormalized couplings are on the same 
classical trajectories determined by the bulk theory. 
Finally, we discussed the relationship between the present formalism 
and the analysis of Henningson and Skenderis, 
and found an algorithm determining the local part of 
the on-shell action, $\Sloc[G(x),\phi(x)]$, 
from the divergent terms in their calculation.

\section*{Acknowledgements}

The authors would like to thank M.\ Ninomiya, S.\ Ogushi 
and H.\ Sonoda for useful discussions.
The work of M.F. is supported in part by a Grand-in-Aid
for Scientific Research from the Ministry of Education, Science,
Sports and Culture,
and the work of T.S.\ is supported in part
by JSPS Research Fellowships for Young Scientists.

\appendix

\section{Variations of Curvature}

In this appendix, we list the variations of the curvature tensor, 
Ricci tensor and Ricci scalar with respect to the metric.

Our convention is\footnote{
The sign is opposite to that adopted in Ref.\ \cite{HS;weyl}.
}
\begin{eqnarray}
 R^{\mu}_{~\,\nu\lambda\sigma} &\equiv& 
 \partial_{\lambda}\Gamma_{\sigma\nu}^{\mu}
  +\Gamma_{\lambda\rho}^{\mu}\Gamma_{\sigma\nu}^{\rho}
  - (\lambda \leftrightarrow \sigma), \nn
 R_{\mu\nu} &\equiv& R^{\rho}_{~\mu\rho\nu},
 \qquad R\,\equiv\,G^{\mu\nu}\,R_{\mu\nu}. 
\end{eqnarray}
The fundamental formula is 
\ba
\delta\Gamma^\kappa_{\mu\nu}
=\frac{1}{2}\,G^{\kappa\lambda}
 \,\left(\nabla_{\mu}\,\delta G_{\nu\lambda}
 + \nabla_{\nu}\,\delta G_{\mu\lambda} 
 - \nabla_{\lambda}\,\delta G_{\mu\nu}
 \right),
\ea
from which one can calculate the variations of curvatures:
\begin{eqnarray}
 \delta R^\mu_{~\,\nu\lambda\sigma}&=&
  \nabla_\lambda\,\delta\Gamma^\mu_{\sigma\nu}
  -\nabla_\sigma\,\delta\Gamma^\mu_{\lambda\nu},\\
 \delta R_{\mu\nu\lambda\sigma} &=&
  \frac{1}{2}\Bigl[
  \nabla_{\lambda}\nabla_{\nu}\delta G_{\sigma\mu}
  -\nabla_{\lambda}\nabla_{\mu}\delta G_{\sigma\nu} 
  -\nabla_{\sigma}\nabla_{\nu}\delta G_{\lambda\mu}
  +\nabla_{\sigma}\nabla_{\mu}\delta G_{\lambda\nu}\nn
 &&~~ +\delta G_{\mu\rho}\,R^{\rho}_{~\nu\lambda\sigma} 
  -\delta G_{\nu\rho}\,R^{\rho}_{~\mu\lambda\sigma} 
  \Bigr],\\
 \delta R_{\mu\nu} &=& \frac{1}{2}\left[\nabla^{\rho}
  \left(\nabla_{\mu}\delta G_{\nu\rho} + \nabla_{\nu}\delta 
  G_{\mu\rho}\right) - \nabla^2 \delta G_{\mu\nu} - 
  \nabla_{\mu}\nabla_{\nu}
   \left(G^{\rho\lambda}\delta G_{\rho\lambda}\right)\right], \nonumber \\
 \\
 \delta R &=& -\delta G_{\mu\nu}\,
  R^{\mu\nu}+\nabla^{\mu}\nabla^{\nu}
  \delta G_{\mu\nu} - \nabla^2\left(G^{\mu\nu}\delta G_{\mu\nu}\right).
\end{eqnarray}
Here note that 
\ba
 \Bigl[\nabla_\mu,\nabla_\nu\Bigr]\,\delta G_{\lambda\sigma}
  =-\delta G_{\rho\sigma}\,R^\rho_{~\lambda\mu\nu}
  -\delta G_{\lambda\rho}\,R^\rho_{~\sigma\mu\nu}.
\ea

%
\section{Variations of $S_{\rm loc}[G(x),\phi(x)]$}
\setcounter{equation}{0}

In this appendix, we list the variations of $S_{\rm loc}[G(x),\phi(x)]$.

\noindent\underline{\bf Pure gravity}:

If we only consider terms with weight $w\leq4$ of the form 
\begin{equation}
 S_{\rm loc}[G] = \int d^d x\sqrt{G}\left(
 W - \Phi R + XR^2 + YR_{\mu\nu}R^{\mu\nu} 
 + ZR_{\mu\nu\rho\lambda}R^{\mu\nu\rho\lambda}
\right),
\end{equation}
then we have 
\begin{eqnarray}
\frac{1}{\sqrt{G}}\frac{\delta S_{\rm loc}}{\delta G_{\mu\nu}}
 \!\!&=&\!\!\frac{1}{2}\Bigl(
 W - \Phi R + XR^2 + YR_{\mu\nu}R^{\mu\nu} 
 + ZR_{\mu\nu\rho\lambda}R^{\mu\nu\rho\lambda} \Bigr)
  G^{\mu\nu} \nonumber \\
 &&+\,\Phi R^{\mu\nu} -2X \Bigl(RR^{\mu\nu}-\nabla^{\mu}
 \nabla^{\nu}R\Bigr) 
 -Y\Bigl(2R^{\mu}_{\,~\rho}R^{\nu\rho}
 -2\nabla_{\rho}\nabla^{\left(\mu\right.}R^{\left.\nu\right)\rho}
 +\nabla^2R^{\mu\nu}\Bigr) 
 \nonumber \\
 && -2Z\Bigl(R^{\mu}_{~\,\rho\lambda\sigma}
 R^{\nu\rho\lambda\sigma}-2\nabla^{\rho}\nabla^{\lambda}
 R^{\left(\mu~~\nu\right)}_{~~\rho\lambda~}\Bigr)
 -\left(2X+\frac{1}{2}\,Y\right)G^{\mu\nu}\,\nabla^2R, 
\end{eqnarray}
and thus
\begin{eqnarray}
 \frac{1}{\sqrt{G}}\,G_{\mu\nu}\,
  \frac{\delta S_{\rm loc}}{\delta G_{\mu\nu}}
 &=&\frac{d}{2}\,W - \frac{d-2}{2}\,\Phi\,R 
 +\frac{d-4}{2}\left(XR^2+YR_{\mu\nu}R^{\mu\nu}+ZR_{\mu\nu\rho\lambda}
 R^{\mu\nu\rho\lambda}\right) \nonumber \\
 &&~- \left(2(d-1)X+\frac{d}{2}\,Y+2Z\right)\nabla^2R.
\end{eqnarray}
In the last expression, we have used the Bianchi identity: 
$\nabla^\mu R_{\mu\nu}=(1/2)\nabla_\nu R$.

\noindent\underline{\bf Gravity coupled to scalars}:

{}For $S_{\rm loc}[G,\phi]$ of the form
\begin{equation}
 S_{\rm loc}[G,\phi] = \int d^d x \sqrt{G}\left(
 W(\phi) - \Phi(\phi)R + \frac{1}{2}M_{ij}(\phi)G^{\mu\nu}
 \partial_{\mu}\phi^i \partial_{\nu}\phi^j 
\right),
\end{equation}
we have
\begin{eqnarray}
\frac{1}{\sqrt{G}}\frac{\delta\Sloc}{\delta G_{\mu\nu}} &=&
 \frac{1}{2}\left(W-\Phi R + \frac{1}{2}\,M_{ij}\,
 \partial_{\rho}\phi^i\,\partial^{\rho}\phi^j \right)G^{\mu\nu} \nn
 &&+\,\Phi \,R^{\mu\nu}
 +G^{\mu\nu}\,\nabla^2\Phi - \nabla^{\mu}
  \nabla^{\nu}\Phi-\frac{1}{2}\,M_{ij}\,
  \partial^{\mu}\phi^i\,\partial^{\nu}\phi^j, \\
\frac{1}{\sqrt{G}}\frac{\delta\Sloc}{\delta \phi^i} &=&
 \partial_i W - \partial_i \Phi\, R -M_{ij}\,\nabla^2\phi^j
 - \Gamma_{i;jk}^{(M)}\,\partial_{\mu}\phi^j\,\partial^{\mu}\phi^k,
\end{eqnarray}
where $\Gamma_{jk}^{(M)i}(\phi)\equiv
M^{il}(\phi)\,\Gamma_{l;jk}^{(M)}(\phi)$ is the Christoffel symbol 
constructed from $M_{ij}(\phi)$.

%
\section{RG Flow and the Classical Solutions in the Bulk}
\setcounter{equation}{0}

According to the holographic RG, 
the RG flow in the boundary field theory should be described 
by the classical solutions in the bulk. 
Although this is clearly explained for $d=4$ in Ref.\ \cite{dVV}, 
we repeat their argument for arbitrary dimensions, 
in order to make our discussion self-contained. 
To this end, we start with the classical solutions 
$\Gb_{\mu\nu}(x,r;G(x),r_0)$ and $\phib^i(x,r;\phi(x),r_0)$ with 
the boundary conditions 
\begin{equation}
 \Gb_{\mu\nu}(x,r_0)=G_{\mu\nu}(x)\equiv{1\over a^2}\,\delta_{\mu\nu},
 \qquad
 \phib^i(x,r_0)=\phi^i(x)\equiv\phi^i={\rm const.} 
\end{equation}
Since we set the fields to constant values, the system 
is now perturbed finitely. 
{}Furthermore, since $a$ gives the unit length of the metric 
$G_{\mu\nu}(x)$, 
this perturbation should describe the system 
with the cutoff length $a$, which corresponds to the time 
$r=r_0$ in the RG flow.
From Eq.\ (\ref{Pi;gphi}) and the Hamilton-Jacobi equation 
(\ref{momentum;hj}), we obtain 
\begin{eqnarray}
 {d\over dr}\Gb_{\mu\nu}(x,r;G,r_0)\Bigg|_{r=r_0}
  &=&{1\over d-1}\,W(\phi)\,{1\over a^2}\,\delta_{\mu\nu},
  \label{barG} \\
 {d\over dr}\phib^i(x,r;\phi,r_0)\Big|_{r=r_0}
  &=&-L^{ij}(\phi)\,\partial_jW(\phi). 
\label{barphi}
\end{eqnarray}
We then assume that the classical solutions take the following form 
for general $r$: 
\begin{equation}
 \Gb_{\mu\nu}(x,r;G,r_0)={1\over a(r)^2}\delta_{\mu\nu}, \qquad
 \phib^i(x,r;\phi,r_0)=\phi^i(a(r)),
\end{equation}
with $a(r_0)=a$. Note that $a(r)$ can be identified with the cutoff 
length at $r$. It then follows from (\ref{barG}) and (\ref{barphi}) that 
\begin{eqnarray}
 a\,{ dr \over da}&=&-\,{2(d-1)\over W(\phi)}, \\
 a\,{d \over da}\,\phi^i(a)&=&{2(d-1)\over W(\phi)}\,
 L^{ij}(\phi)\,\partial_jW(\phi). 
\end{eqnarray}
The latter agrees with the beta function in Eq.\ 
(\ref{cs;beta}).

%
\section{Analysis of the Weyl Anomaly \`{a} la 
Henningson and Skenderis}
\setcounter{equation}{0}

It is convenient to introduce the coordinate $\rho\equiv e^{2r}$ 
and rewrite the metric in the following way, as in Ref.\ \cite{HS;weyl}:
\begin{equation}
 ds^2={d\rho^2\over 4\rho^2}+{g_{\mu\nu}(x,\rho) \over \rho}\,
 dx^{\mu}dx^{\nu}.
\label{metric;hs}
\end{equation}
The metric $g_{\mu\nu}(x,\rho)$ is related to our metric,
$G_{\mu\nu}(x,r)$, as
\begin{equation}
 G_{\mu\nu}(x,r)={g_{\mu\nu}(x,\rho) \over \rho}
 \qquad \left(\,\rho=e^{2r}\,\right).
\label{trans;metric}
\end{equation}
Assuming the existence of an asymptotically AdS${}_{d+1}$ 
boundary at $\rho=0$, 
we expand the metric as\footnote{
The logarithmic term always needs to be added at order $d/2$ 
when $d$ is even.
}
\begin{equation}
 g(x,\rho)=g_{(0)}(x)+\rho\,g_{(2)}(x)+\cdots 
  +\rho^{d/2}\left(g_{(d)}(x)+\log\rho~ h_{(d)}(x)\right)
  +O(\rho^{d/2+1}).
\end{equation}
Then the equations of motion for $g_{\mu\nu}$,
\begin{eqnarray}
 0&=&\tr(g^{-1}g'')-\frac{1}{2}\tr(g^{-1}g'g^{-1}g'),\\
 0&=&\nabla^{\nu}g^{\prime}_{\mu\nu}-\partial_{\mu}
  \tr(g^{-1}g^{\prime}),\\
 0&=&-{\rm Ric}(g)+\rho\left[ 2g^{\prime\prime}
  -2g^{\prime}g^{-1}g^{\prime}
  +\tr (g^{-1}g^{\prime})g^{\prime} \right]
  -(d-2)g^{\prime}-\tr (g^{-1}g^{\prime})g, \nonumber \\
\end{eqnarray}
can be solved iteratively for small $\rho$, 
giving the coefficient functions $g_{(2)},g_{(4)},\cdots$ 
as functions of $g_{(0)}$ \cite{HS;weyl} (see also Ref.\ \cite{dSS}). 
Here $\nabla_{\mu}$ is the covariant derivative with respect to 
$g_{\mu\nu}$, and 
the prime represents $\partial/\partial \rho$. 
The tensors $g_{(k)}~(k=0,2,\cdots,d-2)$ and $h_{(d)}$ are obtained 
as covariant expressions with respect to $g_{(0)}$. 
Although  $\tr \left(g_{(0)}^{-1}g_{(d)} \right)$ is an invariant scalar, 
$g_{(d)}$ itself cannot be expressed covariantly. 
The quantity $\tr \left(g_{(0)}^{-1}h_{(d)} \right)$ turns out to 
vanish identically. 
Then, substituting the classical solution into the bulk action, 
we can explicitly evaluate the dependence of the on-shell action 
on the coordinate of the boundary, $\rho\equiv\epsilon$:
\begin{equation}
 S^{\rm HS}[g_{(0)},\epsilon]=
 \int d^dx\left[\,d \int_{\epsilon}d\rho\,\sqrt{g}\,\rho^{-d/2-1}
 +\left( 4\rho^{-d/2+1}\sqrt{g}^{\,\prime}-2\,d\,\rho^{-d/2}\sqrt{g}
 \right)\Bigg|_{\rho=\epsilon}\right]. 
\end{equation}

\noindent\underline{\bf $d=4$}:

The coefficients necessary for the calculation 
are (using the convention described in Appendix A) 
\begin{eqnarray}
 g_{(2)} &=&
-{1\over 2}\left( {\rm Ric}(g_{(0)})-{1\over 6}R(g_{(0)})g_{(0)}
  \right), \\
 \tr\left[g_{(0)}^{-1}g_{(4)}\right] &=& {1\over 16}
  \left( R^{\mu\nu}(g_{(0)})R_{\mu\nu}(g_{(0)})
   -{2\over 9}(R(g_{(0)}))^2\right), \\
 4h_{(4)} &=& 2g_{(2)}g_{(0)}^{-1}g_{(2)}
+{\rm Ric}^{\prime}(g)|_{\rho=0}
  +\tr \left( 2g_{(0)}^{-1}g_{(4)}-g_{(0)}^{-1}g_{(2)}g_{(0)}^{-1}g_{(2)} 
  \right)g_{(0)}. \nonumber \\ 
\end{eqnarray}
The on-shell action is thus evaluated as 
\begin{eqnarray}
 S^{\rm HS}[g_{(0)},\epsilon]=\int d^4x\sqrt{g_{(0)}}\left(
  -{6\over \epsilon^2}
  +{3\over\epsilon}R(g_{(0)})
  -\log\epsilon\,\,\cW_4[g_{(0)}]\right)
  + \Gamma^{\rm HS}_{\rm fin}[g_{(0)},\epsilon]. \nonumber \\ 
\label{sloc4;hs}
\end{eqnarray}
Here $\cW_d[g_{(0)}]$ is the $d$-dimensional Weyl anomaly 
written in terms of $g_{(0)}$, 
and $\Gamma^{\rm HS}_{\rm fin}[g_{(0)},\epsilon]$ is 
the finite part in the limit $\epsilon\rightarrow0$.

\noindent\underline{\bf $d=6$}:

The calculation is completely parallel to that for the $d=4$ case, 
and we find
\begin{eqnarray}
g_{(2)} &=& -\frac{1}{4}\left({\rm Ric}(g_{(0)})
- \frac{1}{10}Rg_{(0)}\right), \\
{\rm tr}\left[g_{(0)}^{-1}g_{(4)}\right] &=& 
 \frac{1}{64}{\rm tr} \left({\rm Ric}(g_{(0)})^2\right) 
 - \frac{7}{3200}R(g_{(0)})^2, \\
{\rm tr}[g_{(0)}^{-1}g_{(2)}g_{(0)}^{-1}g_{(2)}] &=&
 \frac{1}{16}{\rm tr}\left({\rm Ric}(g_{(0)})^2\right)
 -\frac{7}{800}R(g_{(0)})^2, 
\end{eqnarray}
from which we calculate 
\ba
 &&S^{\rm HS}[g_{(0)},\epsilon]=\int d^6x \sqrt{g_{(0)}}\left( 
  -{10 \over \epsilon^3}+{1 \over 4\epsilon^2}R
  +{3 \over 640\epsilon}R^2-{1\over 64\epsilon}R_{\mu\nu}R^{\mu\nu}
  -\log\epsilon\,\,\cW_6[g_{(0)}]\right)\nn
 &&~~~~~~~~~~~~~~~~~~~~+ \Gamma^{\rm HS}_{\rm fin}[g_{(0)},\epsilon].
\ea

Since the metric appears in the bulk action 
only through the combination $G_{\mu\nu}(x,r)=g_{\mu\nu}(x,\rho)/\rho$, 
we obtain the relation 
\begin{equation}
 S^{\rm HS}[e^{2\sigma}g_{(0)},e^{2\sigma}\epsilon]
  =S^{\rm HS}[g_{(0)},\epsilon],\label{scaling}
\end{equation}
which implies that the coefficient of $\log\epsilon$ 
actually gives the anomaly 
\ba
  \Gamma^{\rm HS}_R[e^{2\sigma} g_{(0)}]
  -\Gamma^{\rm HS}_R[g_{(0)}]
  =2\int d^d x\sqrt{g_{(0)}}\,\,\cW_d[g_{(0)}]\,\sigma ~~~(\sigma \ll 1).
\ea 
where $\Gamma^{\rm HS}_R[g_{(0)}]\equiv 
\lim_{\epsilon\rightarrow 0} 
\Gamma^{\rm HS}_{\rm fin}[g_{(0)},\epsilon]$.
Note also that Eq.\ \eq{scaling} implies that 
$S^{\rm HS}[g_{(0)},\epsilon]$ depends only on $g_{(0)}/\epsilon$.

\end{document}